\documentclass{emulateapj}
\usepackage{apjfonts}
\newcommand{\etal}{{et al.}}
\newcommand{\kms}{km s$^{-1}$}

\newcommand{\tnm}{\tablenotemark}
\newcommand{\tnt}{\tablenotetext}

\begin{document} 

\shorttitle{WLM-1}
\shortauthors{Stephens, Catelan, \& Contreras}
\journalinfo{The Astronomical Journal, 2005 October}
\submitted{Received 2005 Oct 3}

\title{WLM-1: A Non-Rotating, Gravitationally Unperturbed, \\
	Highly Elliptical Extragalactic Globular Cluster}

\author{Andrew W. Stephens\altaffilmark{1}}
\affil{Gemini Observatory, 670 N. A'ohoku Place, Hilo, HI 96720}
\email{stephens@gemini.edu}

\author{M\'arcio Catelan}
\affil{Pontificia Universidad Cat\'olica de Chile, Departamento de 
       Astronom\'\i a y Astrof\'\i sica, \\ Av. Vicu\~{n}a Mackenna 4860, 
       782-0436 Macul, Santiago, Chile}
\email{mcatelan@astro.puc.cl} 

\and

\author{Roxana P. Contreras}
\affil{University of Missouri, Department of Physics \& Astronomy,
503J Benton Hall, 8001 Natural Bridge Road, St. Louis, MO 63121}
\email{rpcontre@astro.puc.cl}

\altaffiltext{1}{Princeton-Cat\'olica Fellow at Princeton University 
Observatory \& Pontificia Universidad Cat\'olica de Chile during part of
this work.}

\begin{abstract} 

Globular clusters have long been known for presenting (at times)
significant deviations from spherical symmetry.  While rotation has been
the main proposed explanation, other complicating factors such as their
constant interaction with the strong gravitational potential of their
host galaxy have made it difficult for a consensus to be reached.  To
address this question we have obtained high-resolution spectra of 
WLM-1, the lone globular cluster associated with the
isolated, low-mass dwarf irregular galaxy WLM. Using archival HST WFPC2
data, we measure the radial ellipticity profile of WLM-1, finding it to
be highly elliptical, with a mean value of 0.17 in the region
$0.5-5''$---which is comparable to what is found in our Galaxy for the
most elliptical globular clusters.  There is no evidence of isophote
twisting, except for the innermost regions of the cluster ($r < 0.5''$).
To investigate whether the observed flattening can be ascribed to
rotation, we have obtained long-slit high-resolution VLT/UVES spectra of
this cluster along and perpendicular to the axis of flattening.  Using
cross-correlation we find that the velocity profile of the cluster is
consistent with zero rotation along either axis.  Thus neither cluster
rotation nor galactic tides can be responsible for the flattened
morphology of WLM-1.  We argue that the required velocity dispersion
anisotropy between the semi-major and semi-minor axes that would be
required to account for the observed flattening is relatively small, of
order 1~\kms.  Even though our errors preclude us from conclusively
establishing that such a difference indeed exists, velocity anisotropy
remains at present the most plausible explanation for the shape of this
cluster.

\end{abstract} 


\section{Introduction} \label{sec:introduction} 

Globular clusters have long been noted to deviate, sometimes
substantially, from perfect spherical symmetry
\citep{Shapley1917,Pease1917}. What is the reason why many globular
clusters are non-spherical?  A conclusive answer to such an important
question in the study of globular cluster dynamics, with potentially
important ramifications in the study of the evolution of globular
cluster stars \citep{Norris1983,Norris1987}, 
has so far not been conclusively established
\citep[see][ for an extensive review and references]{Meylan1997}.

Several possible causes of deviations of globular cluster shapes from
sphericity have been suggested, including tidal stresses due to the
presence of strong galactic tidal fields \citep{Longaretti1997,
Combes1999}, gravothermal shocks during the passage of a cluster through
the disk \citep{Kontizas1989}, velocity anisotropies, and internal
rotation (which may also be acquired by interaction with a massive
galaxy; Lee et al. 2004 and references therein). In particular, the
former two phenomena are now well documented, with a particularly
impressive example being given by Palomar~5 \citep{Odenkirchen2001}. The
possibility that rotation may be the main driver of cluster
ellipticities \citep{Ryden1996} is supported by the \citet{Meylan1986}
and \citet{Merritt1997} studies of $\omega$~Centauri (NGC~5139), which
are based upon a detailed analysis of its structural and dynamical
profiles. Yet, it still remains unclear, both from an empirical and from
a theoretical perspective, whether rotation can be the main cause of
flattening of globular star clusters.

On the observational side, we currently lack detailed velocity profiles
for most globular clusters; and even though we do have detailed
rotational information for a few clusters, such as 47~Tucanae (NGC~104),
$\omega$~Cen, M4 (NGC~6121) and M15 (NGC~7078)
\citep{Mayor1984, Gebhardt1995, Peterson1995, Gerssen2002, Zoccali2004},
the lack of a wider sample, as well as of reliable measurements of the
variation in flattening as a function of radius for clusters
encompassing the full range of observed ellipticities, has hindered
further progress in establishing the general correspondence, or lack
thereof, between flattening and cluster rotation.  Indeed, the latest
catalogs of globular cluster ellipticities, published by
\citet{Frenk1982} and by \citet{White1987}, are entirely based on
photographic data, consider exclusively the outermost cluster regions,
and, relying on visual data, can be severely affected by differential
reddening \citep{vandenBergh1982}.  Our group is currently building a new, 
detailed catalog of globular cluster ellipticities, based on 2MASS 
(near-infrared) data, which will greatly improve on the current 
situation \citep{Navarro2006}.

Theoretically, too, the case for rotation as one of the main drivers of
globular cluster ellipticity is far from being settled.  In particular,
\citet{Meza2002} has recently argued that globular clusters can rotate
to very fast rates without deviating substantially from sphericity.
This result is in contrast with those from \citet{Alimi1999}, who
instead were unable to produce models of fast-rotating globular
cluster-like structures {\em without} an accompanying deviation from
sphericity, in the sense that their modeled structures became lengthened
along one or two axes, depending on the initial anisotropy of the
velocity distributions.


An ideal test, from an empirical viewpoint, of the impact of rotation on
the shapes of globular star clusters, is to measure the rotation
velocity of a highly flattened, isolated extragalactic globular cluster,
where there is little possibility that tidal effects and/or gravothermal
shocks may be responsible for its flattening.  The closest possible
approximation to such an isolated, flattened system is represented by
the single, old globular cluster in the Wolf-Lundmark-Melotte (WLM)
dwarf irregular galaxy (DDO 221). While distant
\citep[885~kpc;][]{Hodge1999}, the mass of the WLM galaxy is much smaller
than even the LMC or SMC, with $M_V$ of $-14.4$, $-18.5$, and $-17.1$,
respectively (van den Bergh 2000), thus effectively ruling out the
possibility that the non-spherical shape of the WLM cluster (hereafter
WLM-1) can be due to tidal effects or gravothermal shocks. Note, in this
sense, that this is not necessarily the case even in the Magellanic
Clouds \citep{Goodwin1997}, where non-spherical globular clusters of all
ages can be found \citep[][and references therein]{vandenBergh1991}.

In this paper we begin by reanalyzing archival HST WFPC2 data in
Section~\ref{sec:hstobservations}, where we measure the ellipticity and
surface brightness profiles of WLM-1.  We then discuss our new VLT UVES
observations and their reduction in Section~\ref{sec:vltobservations},
and attempt to measure the rotation (Section~\ref{sec:rotation}) and
velocity dispersion (Section~\ref{sec:veldisp}) of WLM-1.  We conclude
with a discussion of the possibility of velocity anisotropy in WLM-1 as
a possible explanation for its observed morphology, and give some final
remarks in Section~\ref{sec:conclusions}.

\section{HST WFPC2 Observations} \label{sec:hstobservations} 

\begin{figure}[htb]
\epsscale{1.1}
\plotone{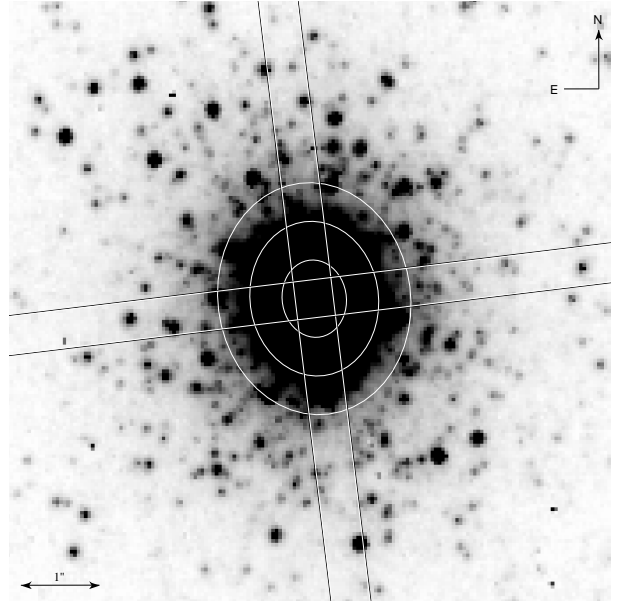} 
\figcaption{
A combined F555W/F814W HST image of WLM-1 \citep{Hodge1999}.  The
approximate size and orientation of the UVES slits are indicated.  We
have also inscribed ellipses illustrating the mean measured ellipticity
($0.17$) and position angle ($7.1\degr$), and the narrow extraction
limits.  The size of the image is approximately $7.7''$, north is up and
east is left.
\label{fig:wlm-1}}
\end{figure}

In order to measure accurate ellipticity and surface brightness profiles
of WLM-1, we have downloaded calibrated HST WFPC2 F814W and F555W images
from the HST archive (program GO~6813, PI P. W. Hodge).  In these images
WLM-1 is centered on the PC chip.  The only processing required was to
combine the two images in each filter to reject cosmic rays.  We also
combined the images from different filters to reject the remaining few
cosmic rays.  This final image, rotated to put north at the top, is
shown in Figure~\ref{fig:wlm-1}.

\subsection{Cluster Ellipticity} \label{sec:ellipticity} 

\begin{figure}[htb]
\epsscale{1.1}
\plotone{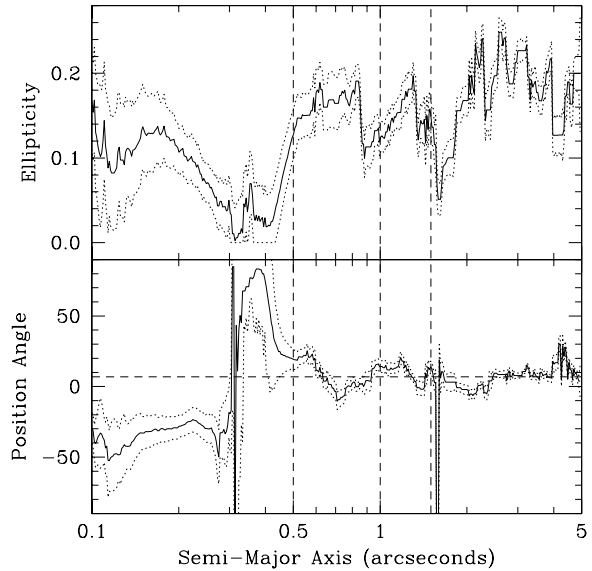} 
\figcaption{
Ellipticity and position angle of WLM-1 as a function of semi-major axis
distance as measured on the combined F555W and F814W frame.  Dotted
lines show $1 \sigma$ measurement errors.  Vertical dashed lines at 0.5,
1.0, and $1.5''$ illustrate our spectral extraction limits.  The
horizontal dashed line indicates the major-axis slit position angle.
\label{fig:ellipticity}}
\end{figure}

The ellipticity and position angle as a function of semi-major axis
distance are measured with the Image Reduction and Analysis Facility
(IRAF\footnote{IRAF is distributed by the National Optical Astronomy
Observatories, which are operated by AURA, Inc., under cooperative
agreement with the NSF.}) routine {\sc ellipse}, where $\epsilon = 1 - b/a$,
and $a$ and $b$ are the semi-major and semi-minor axis lengths
respectively.  The position angle is the angle
($-90\degr<\alpha<90\degr$) made by the major axis measured degrees east
of north.  The results are illustrated in Figure~\ref{fig:ellipticity}.
The mean ellipticities and position angles for our spectral extraction
regions are listed in Table~\ref{tab:ellipticity}.

\begin{deluxetable}{lcc}
\tablewidth{0pt}
\tablecaption{Ellipticity}
\tabletypesize{\footnotesize}
\tablehead{ 
\colhead{$a (\arcsec)$} &
\colhead{$\epsilon$} &
\colhead{${\rm PA} (\degr)$} }
\startdata
0.1 -- 0.5 & $0.074 \pm 0.042$ & $4.7 \pm 44.1$ \\ 
0.5 -- 1.0 & $0.153 \pm 0.025$ & $7.0 \pm 9.8$ \\ 
1.0 -- 1.5 & $0.150 \pm 0.021$ & $9.4 \pm 6.5$ \\ 
1.5 -- 5.0 & $0.180 \pm 0.038$ & $7.7 \pm 6.6$
\enddata
\label{tab:ellipticity}
\end{deluxetable}

Note that since there are many resolved stars in the HST image, there
are many irregularities in the measurement of the ellipticity and
position angle, especially at larger radii where these stars contribute
a significant fraction of the total flux (see discussion in Navarro et
al. 2005).  Also, when the ellipticity gets close to zero (nearly
circular), the position angle becomes poorly constrained and shows large
fluctuations.


The mean ellipticity of WLM-1 in the region $0.5'' < r < 5''$ is $0.172
\pm 0.039$.  This ranks among the highest values among Galactic
globulars according to the lists tabulated by \citet{Frenk1982} and
\citet{White1987}, and only $\sim10$ Galactic globular clusters have
ellipticities greater than this according to the recent analysis of
2MASS data by Navarro et al. (2005). It has previously been noted that
LMC and SMC globular clusters of all ages also have significant
ellipticities, and in fact tend to be more elliptical than either
Galactic or M31 clusters \citep[][and references
therein]{vandenBergh1991, Goodwin1997}, suggesting an inverse
relationship between galaxy mass and globular cluster ellipticity which
would appear to be broadly consistent with the case of WLM-1. However,
\citet{Harris2002} have recently shown that the giant elliptical galaxy
NGC~5128, with $M_V = -20.1$~mag \citep{deVaucouleurs1980}, also
contains a sizable population of flattened clusters.

It is also interesting to note that \citet{Han1994} have suggested that,
while Galactic and M31 globular clusters shapes are broadly consistent
with oblate spheroids, LMC and SMC clusters are instead more consistent
with triaxial ellipsoids. As well known, triaxiality is often
accompanied by isophote twisting \citep{Williams1979}. In the case of
WLM-1, we do find some interesting evidence that isophote twisting may
indeed be present, but this is clearly restricted to the innermost
$0.5''$ of the cluster (Fig.~\ref{fig:ellipticity}).

\subsection{Surface Brightness Profile} \label{sec:sbprofile} 

\begin{figure}[t]
\epsscale{1.1}
\plotone{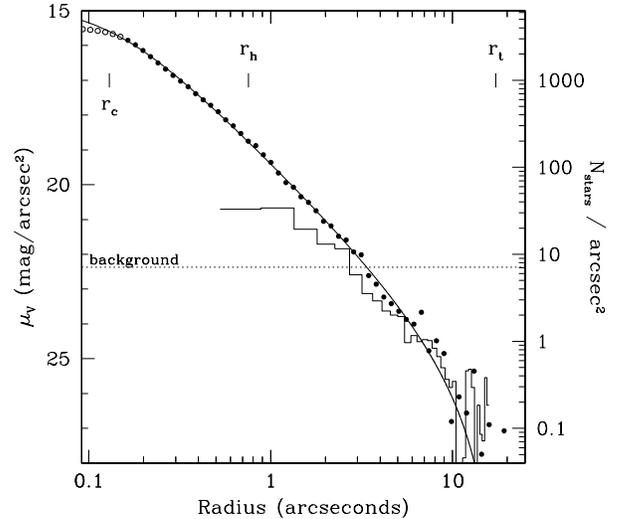} 
\figcaption{
The background-subtracted $V$-band surface brightness profile of WLM-1
(points) and the best-fit King model (line) as determined via
chi-squared minimization ignoring the saturated measurements (open
points).  The fit parameters, core radius ($r_c$) and tidal radius
($r_t$), are marked at the top, as well as the background level (dotted
line) and the half-light radius ($r_h$).  The background-subtracted star
count data from \citet{Hodge1999} have also been overplotted for
comparison (histogram), with units shown on the right-hand axis.
\label{fig:sbprofile}}
\end{figure}

We have independently measured the surface brightness profile on HST
WFPC2 F555W ($\sim V$) and F814W ($\sim I$) images obtained from the
HST archive.  Using the IRAF {\sc ellipse} task, with the ellipticity
and position angle fixed at the mean values of $0.17$ and
$7.1\degr$ respectively, we measured the average flux in each elliptical
annulus, which were then converted to surface brightness using the
photometric zero points (ZP) from the HST manual [${\rm ZP(555)} =
22.545$, ${\rm ZP(814)} = 21.639$].  The data from the F555W filter are
shown as points in Figure~\ref{fig:sbprofile}.

%
%
%

Finally we fit the surface brightness measurements with \citet{King1962}
profiles using a chi-squared minimization algorithm and using the RMS
scatter in each annulus as an estimate of the measurement errors
(although this is actually a measure of the true range of flux values at
any particular semi-major axis distance rather than a measure of how
accurately one can measure the surface brightness).  Pixels within $\sim
0.15''$ of the center of the cluster are saturated, and were therefore
excluded from the fit.  These excluded points are indicated by open
circles in Figure~\ref{fig:sbprofile}.  The resulting fits for each
filter were then averaged together to give the final values.  We find a
best-fit core radius of $r_c = 0.126'' \pm 0.021''$, and a tidal radius
of $r_t = 16.3''\pm 10.4''$, where the magnitude of the errors is driven
by the large scatter of fluxes in each annulus (due to the resolved
nature of the cluster population).  The fitting also takes into account
the background surface brightness; the best fit values are $22.37 \pm
0.13$ mag/arcsec$^2$ in F555W and $21.45 \pm 0.13$ mag/arcsec$^2$ in the
F814W filter. Note that we have {\em not} taken into account the finite
size of the HST PSF, which has a nominal value of less than $0.1''$ in
both filters.


Assuming $(m-M)_0 = 24.73$ \citep{Hodge1999}, one arcsecond corresponds
to 4.28 parsecs at the distance of WLM-1, and our measurements of the
core and tidal radii convert to $r_c = 0.54 \pm 0.09$~pc and $r_t = 70
\pm 44$~pc.  Both of these values fall squarely in the range of
parameters exhibited by Galactic globulars: $0.03 < r_c({\rm pc}) <
21.9$ and $1.6 < r_t({\rm pc}) < 214$ with median values of $r_c =
1.06$~pc and $r_t = 35.4$~pc \citep{Harris1996}. This implies a very
compact globular cluster, with $c \simeq 2.1 \pm 0.3$.  We cannot
discard the possibility of a collapsed core.  Note that our measured
half-light radius for the WLM-1 cluster, along with its $M_V = -8.74$
(Sec.~\ref{sec:veldispest}), are not inconsistent with the trend between
$r_h$ and $M_V$ for globular clusters in dwarf galaxies and in the outer
Galactic halo suggested by \citet[][their Fig.~7]{vandenBergh2004}.




The structural parameters of WLM-1 have been previously derived using
starcounts by \citet{Hodge1999} who found a core radius of $1.09''
\pm 0.14''$ ($4.6 \pm 0.6$ pc) and a tidal radius of $31'' \pm 15''$
($130 \pm 60$ pc). A comparison of the background-subtracted star count
profile and background-subtracted surface brightness profile
(Fig.~\ref{fig:sbprofile}) yields a very good match down to $\sim 2''$,
where the star counts start to become incomplete.  At $\sim 1''$ the two
profiles diverge; the star counts turn over while the surface brightness
profile continues to rise down to $\sim 0.1''$, at which point the CCD
becomes saturated.  We speculate that this incompleteness in the star
counts is the source of Hodge's large parameter values.





Other quantities which will be of use later in this paper include the
half-light radius $r_h$, the central $V$-band surface brightness
$\mu_V(0)$, and the mean surface brightness inside the half-light radius
$\langle\mu_V\rangle_h$.  We estimate these parameters from the best-fit
King model to the F555W ($V$) band HST data.  We find: 
$r_h = 0.75^{+0.27}_{-0.31}$ arcseconds ($3.2^{+1.2}_{-1.3}$~pc),
$\mu_V(0) = 14.83^{+0.32}_{-0.46}$~mag/arcsec$^2$, and
$\langle\mu_V\rangle_h = 17.31^{+0.23}_{-0.50}$~mag/arcsec$^2$.

\section{VLT UVES Observations} \label{sec:vltobservations} 

The new data analyzed in this paper were obtained with the UV-Visual
Echelle Spectrograph (UVES) on the 8.2 meter VLT UT2 (Kueyen) telescope.
UVES is a two-arm cross-dispersed echelle spectrograph with a dichroic
splitting the red and blue components of the light.  The blue light is
recorded on a single 2K $\times$ 4K EEV CCD, and the red light on a
mosaic of an EEV CCD and an MIT/LL CCD for the reddest orders.

In order to obtain the highest spatial resolution we used a $0.5''$ wide
slit, which allows us to obtain velocity information on a similar scale.
While slightly undersampling the spectral information, we chose to bin
the CCDs $2 \times 2$ to improve the signal-to-noise ratio in the faint
outer parts of the cluster.  This binning yields a plate scale of
$0.492''$/pixel in the blue and $0.364''$/pixel in the red.  The blue
slit has a length of $8''$ ($\sim 16$ pixels) and the red slit $11''$
($\sim 30$ pixels).  The wavelength coverage and resolution of this
configuration is listed in Table~\ref{tab:wavelengths}.


WLM-1 (00:01:49.5, $-15$:27:30.7, J2000) was observed for 3060 seconds
in each of two slit positions, one aligned with the major axis (position
angle ${\rm PA}=6.8\degr$), and another aligned along the minor axis
(${\rm PA}=96.8\degr$).  An HST image of the cluster \citep{Hodge1999}
with the slit orientations indicated is shown in Figure~\ref{fig:wlm-1}.

\begin{deluxetable}{ccccc}
\tablewidth{0pt}
\tablecaption{Wavelength Coverage}
\tabletypesize{\footnotesize}
\tablehead{
\colhead{Region}	&
\colhead{$\lambda_1$~(\AA)}	&
\colhead{$\lambda_2$~(\AA)}	&
\colhead{Dispersion\tnm{a}}	&
\colhead{Resolution\tnm{b}}	}
\startdata
Blue  & 3275 & 4509 & 0.034 & 56,700 \\
Red-L & 4610 & 5607 & 0.038 & 67,220 \\
Red-U & 5682 & 6650 & 0.046 & 66,450
\enddata
\tnt{a}{\AA/pixel}
\tnt{b}{2-pixel resolution}
\label{tab:wavelengths}
\end{deluxetable}

\section{UVES Data Reduction} \label{sec:reduction} 

The data reduction was carried out with IRAF. Each of the three UVES
CCDs (Blue, Red-L, Red-U; see Table~\ref{tab:wavelengths} for wavelength
coverages) were reduced independently, beginning with overscan, bias
subtraction, and removal of the scattered inter-order light.  Before
dividing by the flat field, any variations with scales larger than a few
pixels were removed from the illuminated regions by fitting and dividing
by a high-order spline3 fit using the {\sc apflatten} task.  The
resulting ``flattened'' flat field was then normalized to unity and
divided into the science frames.  Lastly we located and interpolated
over cosmic rays and bad pixels in the science frames using the IRAF
median filtering task {\sc crmedian}.

The center of the cluster was traced in all orders, and then extracted
using the apertures discussed in Section~\ref{sec:extractions}.  The
calibration lamp spectrum and flat field were extracted using identical
traces and extraction apertures as the cluster to insure the most
accurate wavelength calibration.  The extracted science spectra were
then divided by the normalized extracted flat field to remove the blaze
function and any wiggles due to absorption or emission features in the
flat field.
Finally we removed any large-scale problems from the extracted spectra,
such as bad columns or cosmic ray strikes, using linear interpolation
from 100 pixels on either side of the blemish.

The wavelength solution was derived from a Thorium-Argon calibration
lamp spectrum.  The solutions have an RMS scatter of $\sim 0.003$~{\AA},
and use $\sim 1150$, $\sim 750$, and $\sim 450$ lines in each
of the Blue, Red-L, and Red-U wavelength regions respectively.  Note
that we rederived the wavelength solution for each extraction.  We
performed one last cleaning of any remaining cosmic rays in the spectra
by fitting by a cubic spline of high enough order to fit the absorption
features, and iteratively replacing $+4\sigma$ and $-6\sigma$ outliers
with the fit.  Finally we trimmed the noise-dominated ends of each
order, and combined the orders into a single spectrum

\subsection{Extractions} \label{sec:extractions}  

The key to measuring a rotation velocity in data such as these, where a
highly concentrated object is barely resolved, is the careful analysis
of the light coming from the {\em edges} of the cluster.  While the
center of the cluster contains most of the light, and hence has the
highest signal-to-noise ratio, the slit of the spectrograph is $0.5''$ wide,
thus we can obtain no velocity information in the central $0.5''$.  The
seeing, which ranged from $0.53''$ to $0.56''$, and the coarseness of
our binned pixels ($0.492''$/pixel in the blue and $0.364''$/pixel in
the red) also blur the spatial resolution.



We must therefore concentrate on the outer regions of the cluster.
However, the cluster is very compact, with a core radius of $0.126'' \pm
0.021''$ (Sec.~\ref{sec:sbprofile}), and the light profile plummets as
one moves away from the center; the cluster light fades into the
background only $\sim 3''$ from the center.

We have extracted five different regions of various sizes and radii in
hopes of obtaining the best measure of the cluster rotation profile.
Since the cluster is fairly elliptical ($e \simeq 0.16$ in this region)
we scale the minor-axis extraction apertures to insure that we
consistently measure the same physical radius of the cluster.  The
extraction regions are listed in Table~\ref{tab:extractions}.  The first
region includes the entire half of the cluster, and is meant to give a
luminosity-weighted rotation estimate.  However, as mentioned before,
due to the width of the spectrograph slit, the innermost regions will
include an equal mix of approaching and receding stars, so the second
extraction region excludes the inner $0.5''$.  The final three
extraction regions are very narrow, and will be used to try to estimate
the rotation profile of the cluster. 

\begin{deluxetable}{ccccc}
\tablewidth{0pt}
\tablecaption{Extraction Regions\tnm{1}}
\tabletypesize{\footnotesize}
\tablehead{ 
\colhead{}		&
\multicolumn{2}{c}{\underline{\hspace{0.9cm} Major \hspace{0.9cm}}} &
\multicolumn{2}{c}{\underline{\hspace{0.9cm} Minor \hspace{0.9cm}}} \\
\colhead{a} 		&
\colhead{Blue}	&
\colhead{Red}	&
\colhead{Blue}	&
\colhead{Red}	}
\startdata
$0.00 - 1.50''$ & 0.00--3.05 & 0.00--4.12 & 0.00--2.56 & 0.00--3.46 \\
$0.50 - 1.50''$ & 1.02--3.05 & 1.37--4.12 & 0.85--2.56 & 1.15--3.46 \\
$0.00 - 0.50''$ & 0.00--1.02 & 0.00--1.37 & 0.00--0.85 & 0.00--1.15 \\
$0.50 - 1.00''$ & 1.02--2.03 & 1.37--2.75 & 0.85--1.71 & 1.15--2.31 \\
$1.00 - 1.50''$ & 2.03--3.05 & 2.75--4.12 & 1.71--2.56 & 2.31--3.46 \\
background      &    none   & 7.20--13.19 &    none    & 6.04--11.10
\enddata
\tnt{1}{Extraction regions in pixels.}
\label{tab:extractions}
\end{deluxetable}

\begin{figure}[htb]
\epsscale{1.1}
\plotone{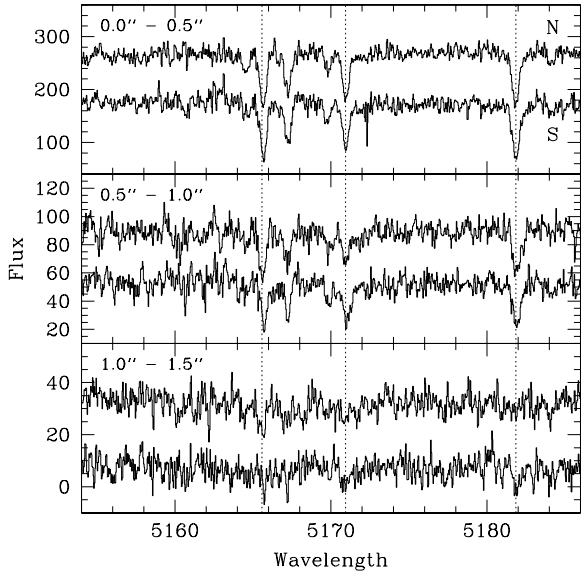} 
\figcaption{
Major axis spectra in three different extractions, showing the 
signal-to-noise level in each.  The two spectra in each panel are from the
north and south extractions (indicated by ``N'' and ``S'') along the
major axis, separated by an equal vertical offset so as not to overlap
in this diagram.  Any significant rotation would be apparent as a
relative shift between these two extractions (a 6.5~\kms\ rotation would
produce a 13~\kms\ shift between the two spectra, corresponding to a 0.22
{\AA} shift at 5170 {\AA}).  The strong features are Mg~I lines
(indicated by the vertical dotted lines).
\label{fig:spectra}}
\end{figure}

When possible the extractions use background subtraction, where the
background is calculated as the median in two regions on either side of
the cluster.  These background regions are scaled according to the
ellipticity for the minor axes extractions.  Note that the slit is too
narrow for background subtraction on the blue chip; however, as there
are no significant sky lines in the blue, this should not affect the
results.

One should note that the WLM-1 cluster is relatively large.  The tidal
radius is $16'' \pm 10''$ (Sec.~\ref{sec:sbprofile}), while the slits
used in this paper to obtain the spectra are only $8''$ and $11''$ long.
Thus the {\it entire} slit has contributions from cluster stars, and any
attempt at background subtraction will also subtract cluster light.

We plot the three narrow spectral extractions along the major axis in
Figure~\ref{fig:spectra}.  This very small piece of the total spectrum
illustrates the variation in signal-to-noise ratio with each extraction
and the abundance of absorption features available for measuring
velocities.


\subsection{Atmospheric Absorption} \label{sec:telluric} 

Terrestrial atmospheric absorption lines are noticeable in our Red-U
spectra.  Two bands are especially strong: H$_2$O at 5880--6020 {\AA},
and O$_2$ at 6200--6360 {\AA}.  We did not observe atmospheric
standards, and therefore we have no way to accurately correct for this
absorption.  However, if not removed these absorption lines will bias
our Red-U derived rotation velocity toward zero since they are at
exactly the same wavelengths on both sides of the cluster.

\begin{figure}[htb]
\epsscale{1.1}
\plotone{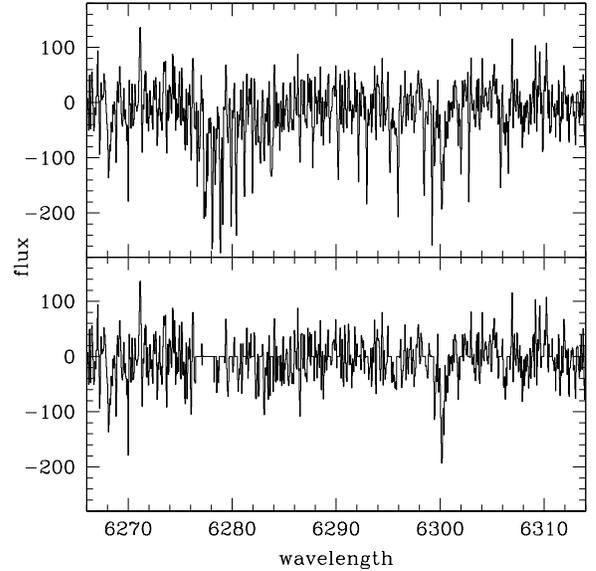} 
\figcaption{
Continuum-subtracted WLM-1 major axis spectrum before and after telluric
masking in the wavelength region of strong atmospheric O$_2$ absorption.
\label{fig:telluric}}
\end{figure}

To remove the telluric absorption lines we simply set any such
absorption to the continuum value.  We start with a high-resolution
atmospheric transmission atlas,\footnote{
ftp.noao.edu/catalogs/atmospheric\_transmission} smooth it to our
resolution, and create a ``telluric mask'' for our spectra.  This mask
has regions of high transmission set to one, and absorption regions,
where the transmission is less than 95\%, are set to a very large
number.  Before cross-correlating the spectra, but after continuum
subtraction, we divide by this telluric mask, and effectively set all
significant regions of atmospheric absorption to the continuum level.
We thus lose any velocity information about WLM-1 which may have been
contained in these masked regions, but the cross-correlation will no
longer be biased by these strong atmospheric features.  A region of
telluric absorption is shown in Figure~\ref{fig:telluric} before and
after masking.

\section{Cluster Rotation} \label{sec:rotation} 

The goal of these observations is to unambiguously measure the rotation
velocity of WLM-1.  One method of getting an estimate of the magnitude
of this rotation would be to follow the original theoretical
calculations of \citet{King1961} for the ellipticity of a rotating
cluster, which is equivalent to assuming that the cluster is a rotating
fluid body.  Assuming a distance modulus of $(m-M)_0 = 24.73$
\citep{Hodge1999}, which corresponds to 4.28 pc/arcsec, we find that
the rotation velocity of WLM-1 should be of the order
\begin{displaymath} \nonumber
v_{\rm rot} = 6.50\, {\rm km s^{-1}} \left( \frac{r}{1''} \right)
	\left( \frac{\rho'}{10^3\, M_{\odot}/{\rm pc}^3} \right) ^{1/2}
	\left( \frac{e}{0.16} \right) ^{1/2},
\end{displaymath}
\noindent where $\rho'$ is the effective density calculated for 
non-homogeneous clusters \citep{vanWijk1949}.  Thus if this cluster is
indeed elliptical as a result of rotation, we might expect to measure a
difference of $\sim 13$~\kms\ along the major axis, and zero along the
minor axis (see Sec.~\ref{sec:veldisp} for a more detailed estimate of
the expected rotation).

While our spectra are not of high signal-to-noise ratio, especially in the
very narrow extractions, we do have a reasonably large wavelength
coverage.  Thus while measuring the position of a single line yields a
relatively large error, combining the results for the hundreds of lines
present across the spectra increases the precision greatly.  One way to
simultaneously compare the positions of all features in two spectra is
cross-correlation.

\begin{figure}[htb]
\epsscale{1.1}
\plotone{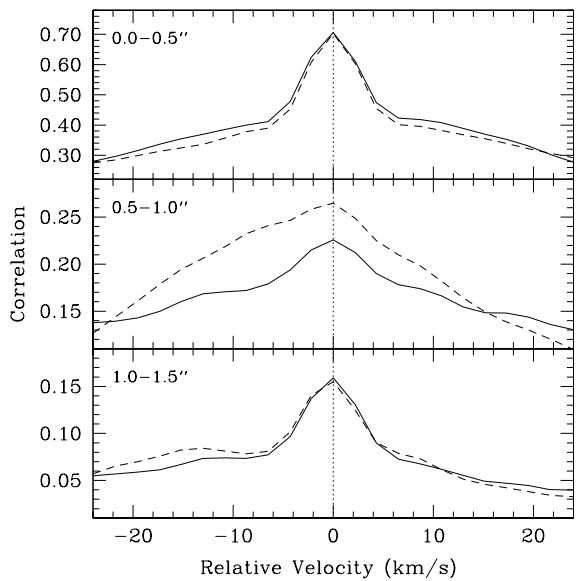} 
\figcaption{
Cross-correlation functions measured across the major (solid) and minor
(dashed) axes, for each of the three extractions: $0-0.25''$ (top),
$0.25-0.50''$ (middle), and $0.50-75''$ (bottom).
\label{fig:xcor}}
\end{figure}

We use the Fourier cross-correlation technique of \citet{Tonry1979},
which has been implemented in the IRAF {\sc fxcor} task.  This routine
computes the height, width, and location of the cross-correlation peak
of two spectra.  For each axis we correlate the north/south (major) and
east/west (minor) extractions which lie at equal radii.  The results of
these correlations for the narrow extractions along both the major and
minor axes are shown in Figure~\ref{fig:xcor}.  The center of the peak
is determined by fitting the top 5 points with Gaussian functions.  This
yields a direct measure of the velocity shift between the two, which is
equal to twice the cluster rotation velocity at that distance.

\begin{figure}[htb]
\epsscale{1.1}
\plotone{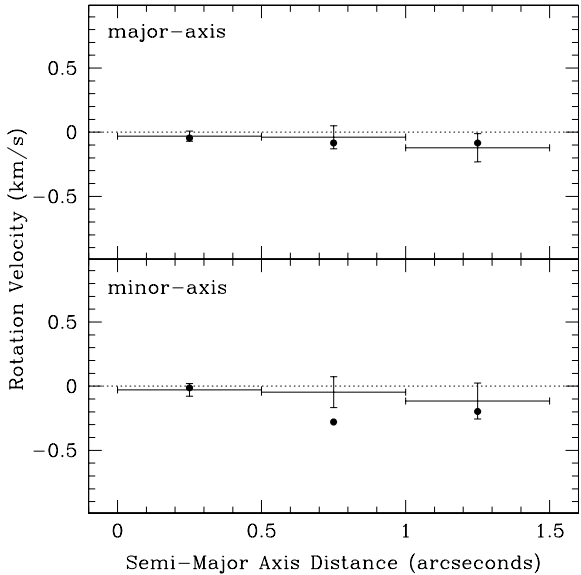} 
\figcaption{
Measured cluster rotation along the major (top) and minor (bottom) axes.
The points are taken from the last three lines of Table~\ref{tab:xcor}
(divided by two to convert to rotation velocity).  The data shown by the
error bars are the results of simultaneously solving for the best
rotation velocities using all 15 narrow extraction cross-correlation
permutations (Table~\ref{tab:vrot}).
\label{fig:rotation}}
\end{figure}

We start with the reduced, extracted spectra from each chip.  In
preparation for the cross-correlation we subtract the continuum by
fitting with a cubic spline.  The telluric absorption is removed from
the Red-U spectra following the procedure outlined in
Section~\ref{sec:telluric}.  We then cross-correlate the spectra from
each (Blue, Red-L, Red-U) chip.  Since the reductions, extractions, and
wavelength calibrations are completely separate for each chip, this
should give three independent, although weaker, measures of the 
cross-correlation velocity for each extraction.  The results of this piecewise
cross-correlation are listed in Table~\ref{tab:piecewisexcor} for all
symmetric extractions, and allow one to compare the quality of the
correlation for each chip. We then combine the spectra from all the
chips, and cross-correlate the result, excluding the $\sim 100$~{\AA}
gaps between the chips.  The results of correlating the combined spectra
are listed in Table~\ref{tab:xcor}, and shown in
Figure~\ref{fig:rotation}.

Rather than only make one cross-correlation measurement to obtain the
velocity at each radius, we can also cross-correlate all possible
permutations, and find the velocities which are most consistent with all
correlations.  In other words, cross-correlating the spectrum extracted
from $r_i$ and $-r_i$ will give a measure of $2 v_i$.  However, one can
also correlate spectra from different radii, $r_i$ and $r_j$, to obtain
the relative velocity difference $\Delta v_{ij}$.

Using all six narrow extractions yields 15 correlation permutations.
Thus we have 15 measurements which depend on only three quantities,
assuming that we should find equal but opposite velocities at the same
radius on either side of the cluster.  We solve for the most probable
velocities by minimizing the sum of the squares of the deviation of each
correlation.

Thus instead of basing a rotation velocity on only one measurement of
the cross-correlation of two spectra at equal distances from the cluster
center, we measure 15 correlations between all 6 spectral extractions
and find the three velocities which are in best agreement with all
measurements.  These ``best fit'' rotation velocities are given as a
function of semi-major axis distance in Table~\ref{tab:vrot}.  We find
no statistically significant rotation along the line of sight at any
distance along either axis of WLM-1.

\begin{deluxetable}{cccc}
\tablewidth{0pt}
\tablecaption{Piecewise Cross-Correlation Velocities\tnm{a}}
\tabletypesize{\footnotesize}
\tablehead{ 
\colhead{region}		&
\colhead{$0.0 - 0.5''$}	&
\colhead{$0.5 - 1.0''$}	&
\colhead{$1.0 - 1.5''$}	}
\startdata
MAJOR-AXIS &                  &                  &                  \\
Blue       & $+0.09 \pm 0.16$ & $+1.77 \pm 0.55$ & $-0.39 \pm 1.58$ \\ 
Red-l      & $-0.20 \pm 0.15$ & $-0.28 \pm 0.78$ & $+0.38 \pm 0.57$ \\ 
Red-u      & $+0.02 \pm 0.11$ & $-0.23 \pm 0.73$ & $-0.20 \pm 0.40$ \\ 
Combined   & $-0.09 \pm 0.08$ & $-0.17 \pm 0.41$ & $-0.17 \pm 0.27$ \\ 
MINOR-AXIS &                  &                  &                  \\
Blue       & $-0.02 \pm 0.15$ & $-0.38 \pm 0.52$ & $-1.48 \pm 0.79$ \\ 
Red-L      & $+0.05 \pm 0.16$ & $+0.02 \pm 0.65$ & $-0.06 \pm 0.54$ \\ 
Red-U      & $+0.03 \pm 0.11$ & $-0.96 \pm 0.65$ & $-0.45 \pm 0.49$ \\ 
Combined   & $-0.03 \pm 0.08$ & $-0.56 \pm 0.32$ & $-0.39 \pm 0.31$
\enddata
\tnt{a}{Velocities in \kms.}
\label{tab:piecewisexcor}
\end{deluxetable}

\begin{deluxetable}{ccccc}
\tablewidth{0pt}
\tablecaption{Combined-Spectrum Cross-Correlation}
\tabletypesize{\footnotesize}
\tablehead{ 
\colhead{}		&
\multicolumn{2}{c}{\underline{\hspace{0.9cm} Major \hspace{0.9cm}}} &
\multicolumn{2}{c}{\underline{\hspace{0.9cm} Minor \hspace{0.9cm}}} \\
\colhead{a} 		&
\colhead{height}	&
\colhead{velocity\tnm{a}}	&
\colhead{height}	&
\colhead{velocity}	}
\startdata
$0.00 - 1.50''$ & 0.68 & $+0.01 \pm 0.08$ & 0.67 & $-0.01 \pm 0.08$ \\ 
$0.50 - 1.50''$ & 0.26 & $-0.29 \pm 0.33$ & 0.28 & $-0.42 \pm 0.28$ \\ 
$0.00 - 0.50''$ & 0.70 & $-0.09 \pm 0.08$ & 0.69 & $-0.03 \pm 0.08$ \\ 
$0.50 - 1.00''$ & 0.23 & $-0.17 \pm 0.41$ & 0.27 & $-0.56 \pm 0.32$ \\ 
$1.00 - 1.50''$ & 0.16 & $-0.17 \pm 0.27$ & 0.16 & $-0.39 \pm 0.31$
\enddata
\tnt{a}{Velocities in \kms.}
\label{tab:xcor}
\end{deluxetable}


\begin{deluxetable}{ccc}
\tablewidth{0pt}
\tablecaption{Rotation Velocities$^{\rm a}$}
\tabletypesize{\footnotesize}
\tablehead{ 
\colhead{a}	&
\colhead{Major}	&
\colhead{Minor}	}
\startdata
$0.00 - 0.50''$ & $-0.03 \pm 0.04$ & $-0.03 \pm 0.05$ \\ 
$0.50 - 1.00''$ & $-0.04 \pm 0.09$ & $-0.05 \pm 0.12$ \\ 
$1.00 - 1.50''$ & $-0.12 \pm 0.11$ & $-0.12 \pm 0.14$
\enddata
\tnt{a}{Velocities in \kms.}
\label{tab:vrot}
\end{deluxetable}



\subsection{Rotation Errors} \label{sec:rotationerrors} 

In order to estimate the robustness of our results we have performed
various cross-correlation simulations.  All of these tests start out
with a base spectrum which is the average of the major- and minor-axis
integrated spectra of WLM-1.

\begin{figure}[htb]
\epsscale{1.1}
\plotone{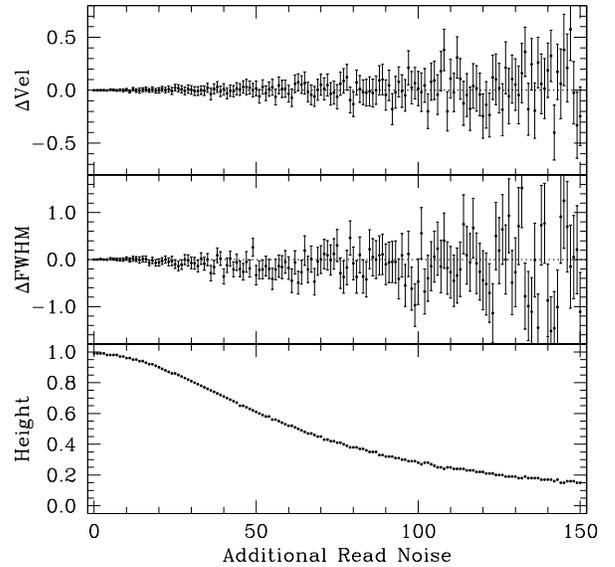} 
\figcaption{
Results of cross-correlation simulations using the combined major- and
minor- axis integrated spectra with varying amounts of additional random
read noise.  The spectra were offset by 5~\kms, and here we illustrate
the difference in the cross-correlation velocity (top), and the width
(middle) and height (bottom) of the cross-correlation peak.  The
velocity errors are those returned by {\sc fxcor}, and the FWHM errors
are three times the velocity error.
\label{fig:sims}}
\end{figure}

To test the dependence of the cross-correlation velocity on the
signal-to-noise ratio of the spectra we create pairs of spectra at
various signal levels by adding random noise to two copies of the base
spectrum, treating them as opposite sides of the cluster.  We then
remove the telluric absorption and shift one of the spectra by 5~\kms.
The pair of spectra are then cross-correlated as described in
Section~\ref{sec:rotation} to obtain their relative velocity.  The
results of these tests are illustrated in Figure~\ref{fig:sims}, along
with the height and FWHM of the cross-correlation peak.  This figure
shows no systematic trends with signal-to-noise ratio, and verifies that the
velocity errors returned by IRAF are indeed valid.  From this analysis
we estimate that the errors associated with the measurement of the 
cross-correlation FWHM are approximately three times the reported velocity
errors, and have plotted them as such in Figure~\ref{fig:sims}.


To estimate the errors of the $\chi^2$-minimization technique used to
determine the ``best'' rotation velocity we ran 50 simulations
consisting of six copies of the integrated spectrum of WLM-1 to mimic
the six $0.5''$ extractions.  Each of the six spectra is shifted such
that the simulated cluster has a monotonically increasing rotation
velocity with radius, and equal but opposite rotation velocities on
opposite sides of the cluster.  Specifically, the spectra were shifted 
by $-3$, $-2$, $-1$, +1, +2, and +3~\kms\ each.  Noise was added to each
spectrum so that the cross-correlation height measured in the simulation
(Fig.~\ref{fig:sims}) matched that measured in the real data (columns
2 and 4 in Table~\ref{tab:xcor}).

The velocity error distributions (recovered -- input) resulting from the
50 simulations have widths of 0.02, 0.05, and 0.06~\kms\ for the central,
middle, and outer velocities respectively.  For each simulation we also
record the average error of each measurement which results from making
all cross-correlation measurements consistent (the square of this
quantity is actually what is minimized when solving for the best set of
rotation velocities).  The mean of this error over all simulations is
0.031~\kms.  However, in the real data, the error is 0.056~\kms\ along
the major axis and 0.074~\kms\ along the minor axis.  This discrepancy
indicates that the simulation was too idealistic, and we have therefore
scaled the estimated velocity errors by 1.8 for the major axis, and 2.4
for the minor axis, yielding our formal error estimates of 0.04, 0.09,
and 0.11~\kms\ for the central, middle, and outer extractions along the
major axis, and 0.05, 0.12, and 0.14~\kms\ along the minor axis.




\section{Cluster Velocity Dispersion} \label{sec:veldisp} 

The velocity dispersion can be derived from the width of the 
cross-correlation peak, which is the sum of the line widths of the two spectra
being correlated added in quadrature.  In the case of the globular
cluster WLM-1, the line width is a combination of the intrinsic stellar
line width, the instrumental profile, and the velocity dispersion of the
cluster.  Unfortunately, no template star was measured at the time of
our observations.

In order to estimate the velocity dispersion of WLM-1, we used two
high-resolution stellar spectra obtained from the UVES Paranal
Observatory Project \citep{Bagnulo2003} as templates.  We chose HD~30562
and HD~45067 (both F8V stars) because of their similarities in line
strengths to the WLM-1 spectrum.  It is important to note that the
template spectra were obtained using the $0.5''$ slit and no binning,
while the WLM-1 spectra used $2 \times 2$ binning.  We therefore
rebinned the template stars to the same resolution as our observations.

In order to increase the signal-to-noise ratio of the WLM-1 spectra, we
averaged together the spectra extracted from opposite sides of the
cluster (since we only want to compare the velocity dispersions along
the major and minor axes).  We then analyzed the spectra measured on
each chip separately to allow optimal Fourier filtering and rejection of
undesirable lines.  For example, we rejected the blue end of the blue
chip which was contaminated by broad hydrogen lines and Ca{\sc ii} H\&K
lines, the region around H$\beta$ on the Red-L chip, and H$\alpha$ on the
Red-U chip.



We then cross-correlated each WLM-1 spectrum with each of the two
template spectra.  This gives a cross-correlation width for each
template, which are then averaged.  The resulting cross-correlation
widths derived for the integrated cluster spectra are 28.7 and 29.0~\kms\
for the major and minor axes respectively.  This width is the quadratic
sum of the WLM-1 and template line widths.  In order to estimate the
line widths of the templates, we cross-correlate the two templates with
one another, and averaging the values from each chip, we find a width of
16.4~\kms.

\begin{figure}[htb]
\epsscale{1.1}
\plotone{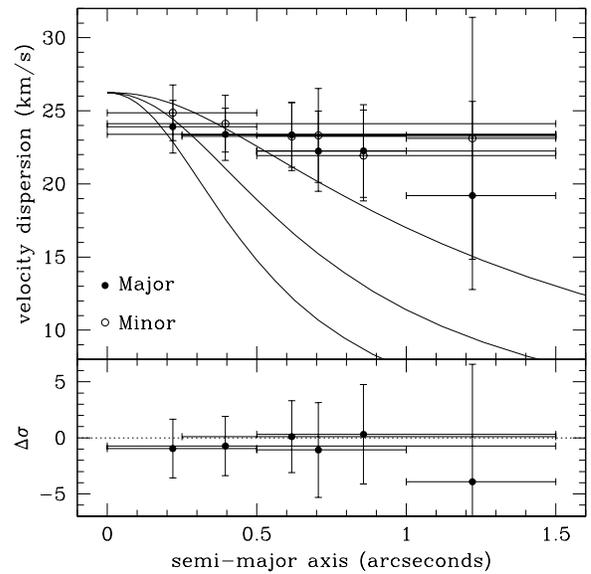} 
\figcaption{
The velocity dispersion estimated via cross-correlation along the
major-axis (filled circles) and minor-axes (open circles) of WLM-1
(top).  The curves show the velocity dispersion profiles predicted by a
King model with $c=$ 2.00, 2.25, and 2.50.  The bottom panel shows the
difference in velocity dispersion between the major and minor axes.
\label{fig:veldisp}}
\end{figure}

If we make the (generous) assumption that the WLM-1 and template spectra
have the same instrumental width and intrinsic stellar line widths, it
is now possible to subtract these components to find the stellar
velocity dispersion.  Subtracting in quadrature yields velocity
dispersions of 23.5 and 23.9~\kms\ along the major and minor axes
respectively.  Following the same procedure for each of the various
spectral extractions of WLM-1 yields the velocity dispersion profile
shown in Figure~\ref{fig:veldisp}, where the errors are three times the
velocity errors, as estimated in Section~\ref{sec:rotationerrors}.  Here
the extraction semi-major axis distances are shown using flux-weighted
centers.

For comparison we have overplotted the velocity dispersion profiles
predicted by the isotropic Maxwellian distribution of the King
model. These curves show the velocity dispersions computed according to
eq. (31) in \citet{King1966} with a concentration $c=$ 2.00, 2.25, and
2.50, scaled to match the observed central velocity dispersion
(but see Sec.~\ref{sec:veldispest}), where
the $c=2.5$ profile is the flattest and, of these three curves, the best
fit to the data. The discrepancy between the model and the measured data
points (given that the light profile indicates $c=2.1 \pm 0.3$) suggests
that either the mass distribution is more extended than predicted by a
King profile or that there is a systematic problem in our measurement of
the velocity dispersion.


In order to verify that any observed differences (or lack thereof) in
line widths are inherent in the source spectra and not due to changes in
the instrumental/atmospheric contribution, we have also compared the
widths of unresolved telluric absorption lines in each using the
atmospheric transmission spectrum described in
Section~\ref{sec:telluric} as the template.  We first excise a region of
each WLM-1 spectrum in the range 6270--6320 {\AA} which is dominated by
absorption lines due to atmospheric O$_2$.  We remove the continuum and
broad features by fitting with a high-order cubic spline, and estimate
the instrumental width of each via cross-correlation with the
atmospheric spectrum.  The widths of the telluric lines present in the
major- and minor-axis spectra differ by less than 0.05~\kms, confirming
the stability of the conditions during the observations.


Our {\em absolute} measurement of the velocity dispersion
profiles is uncertain due to the low signal-to-noise of our spectra, and
the imperfect match with the templates which were obtained at a
different time using a different instrument configuration.  However, the
{\em relative} velocity dispersion between the major and minor axes
should be much more reliable.  The bottom of Figure~\ref{fig:veldisp}
shows the difference between the velocity dispersions measured along the
major and minor axes.  All spectral extraction regions are consistent
with zero difference, and comparing the integrated cluster spectra
measured along the major and minor axes yields a difference of only
$-0.4 \pm 2.6$~\kms\ (major--minor).

\subsection{Central Velocity Dispersion Estimate} \label{sec:veldispest}

We can obtain an estimate of the central velocity dispersion of WLM-1
as a reality check of our measurement of the velocity dispersion
profile.  The simplest way to obtain a rough estimate of the velocity
dispersion is to utilize the correlation between the central velocity
dispersion and the cluster absolute magnitude which was found by
\citet{Djorgovski2003} to be followed by globular clusters in several
different galaxies.  On the basis of Figure~3 (top panel) in Djorgovski
et al., we find

\begin{equation}
\log\sigma = -1.01 - 0.229 \, M_V, 
\end{equation}
 
\noindent where $\sigma$ is in \kms. We use the apparent magnitude of the 
cluster from Table~I in \citet{Sandage1985}, $V = 16.06$~mag, and the
extinction and distance moduli estimated by \citet{Hodge1999}, $A_V =
0.07$~mag and $(m-M)_0 = 24.73$~mag, respectively, to find $M_V =
-8.74$~mag for WLM-1. The above relation then gives a central velocity
dispersion $\sigma \simeq 9.8$~\kms.

A perhaps more accurate estimate for the central velocity dispersion of
the cluster can be obtained from the ``fundamental plane'' correlation
for globular clusters, as provided by \citet{Djorgovski1995}.  In
particular, Djorgovski has shown that (Galactic) globulars follow tight
relations between the central velocity dispersion $\sigma$, the central
surface brightness $\mu_V(0)$, and the core radius $r_c$, on the one
hand; and between $\sigma$, the half-light radius $r_h$, and mean
surface brightness within $r_h$, $\langle \mu_V\rangle_h$.  Based on his
equations~(3) and (6), we find:

\begin{equation}
\log\sigma = 4.173 (\pm 0.175) + 0.45 \, \log r_c - 0.204 (\pm 0.008)\, \mu_V(0), 
\label{eqn:sigrc}
\end{equation}

\noindent and 

\begin{equation}
\log\sigma = 4.829 (\pm 0.237) + 0.70 \, \log r_h - 0.244 (\pm 0.012) \, \langle\mu_V\rangle_h, 
\label{eqn:sigrh}
\end{equation}
 
\noindent where $r_c$ and $r_h$ are in parsecs, and $\sigma$ in \kms. 

Using equation~(\ref{eqn:sigrc}) with the structural parameters derived
in Section~\ref{sec:sbprofile}, core radius $r_c = 0.56$ pc and
central $V$-band surface brightness $\mu_V(0) = 14.83$~mag/arcsec$^2$, 
we estimate the velocity dispersion to be $10.8^{+10.5}_{-5.3}$~\kms.
Equation~(\ref{eqn:sigrh}) with the half-light radius $r_h = 3.21$ pc and
the mean surface brightness inside the half-light radius
$\langle\mu_V\rangle_h = 17.31$ mag/arcsec$^2$ gives a central velocity
dispersion of $9.1^{+16.3}_{-5.8}$~\kms.
These estimates of the central velocity dispersion are lower than our
measurements in Figure~\ref{fig:veldisp}, but the uncertainties do
not exclude our value.  The larger velocity dispersion implies that the
mass of WLM-1 lies on the high side of the relation derived for Galactic
clusters.

\section{Cluster Velocity} \label{sec:velocity} 

Extracting an integrated spectrum from the entire cluster using a $3''$
($\pm 1.5''$) aperture yields a relatively high signal-to-noise spectrum
relative to the partial extractions used to measure cluster rotation.
These integrated spectra have many strong lines from which we can
measure the net radial velocity of the cluster.  In order to make this
measurement, we first calculated and corrected for terrestrial motion,
based on the observation times of 2003/10/03 at 3:23 and 4:23 UT for the
minor and major-axis respectively.  We obtained an atomic line list from
\citet{Reader1980}, and used the IRAF task {\sc rvidlines} to calculate
the radial velocity.  Centroiding $\sim 70$ lines in each of the major
and minor axis spectra yields heliocentric radial velocities of $-105.77
\pm 0.50$ and $-105.84 \pm 0.50$~\kms\ respectively.  Averaging the
results we estimate that the heliocentric radial velocity of WLM-1 is
$-105.8 \pm 0.4$\kms.  The radial velocity of the WLM galaxy is $-116
\pm 2$~\kms\ \citep{Huchra1993}, thus the cluster has a velocity of
$10$~\kms\ with respect to its host.





\section{Velocity Anisotropy}\label{sec:velocityanisotropy} 

It appears that the ellipticity of WLM-1 cannot be due to rotation, so
perhaps WLM-1 is (in a sense) more like elliptical galaxies, where
anisotropies in the velocity dispersion play an important role in the
observed morphology.  To address this hypothesis we employ the tensor
virial theorem, as described in \citet{Binney1987}, to obtain
information about the relation between the ratio of rotation velocity
and velocity dispersion $v_{\rm rot}/\sigma$, on the one hand, and
ellipticity, on the other.

We start by assuming that a globular cluster is an oblate spheroid,
rotating around the $z$ axis, and is seen edge-on. We assume, moreover,
that the system's isodensity surfaces are concentric
ellipsoids. Defining $v_{\rm rot,0}^2$ as the mass-weighted mean-square
rotation speed, and $\sigma_0^2$ as the mass-weighted mean-square random
velocity along the line of sight, we arrive at the following relation:

\begin{equation} 
\frac{v_{\rm rot,0}^2}{\sigma_0^2} = \frac{1-\delta}{\sqrt{1-e^2}}
   \left[\frac{\arcsin(e)-e \sqrt{1-e^2}}{e-\sqrt{1-e^2}\,\arcsin(e)}\right] - 2, 
\end{equation}

\noindent where $\delta < 1$ is a parameter that measures the anisotropy 
of the velocity dispersion tensor and $e$ is the {\em eccentricity},
which is related to the ellipticity $\epsilon$ as follows:
 
\begin{equation}
e = \sqrt{1-(1-\epsilon)^2}. 
\end{equation}

\begin{figure}[t]
\epsscale{1.1}
\plotone{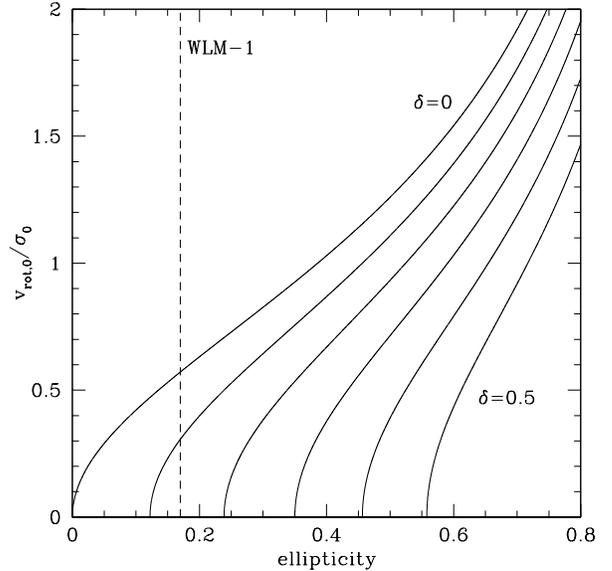} 
\figcaption{
The predicted ratio between mass-weighted rotation velocity and
mass-weighted velocity dispersion is plotted as a function of the
ellipticity for an oblate spheroid. The curves correspond to values of
the velocity anisotropy factor $\delta$, with an interval of 0.1 in
$\delta$ from the upper to the lower curve. The dashed line indicates
the measured ellipticity value for WLM-1.
\label{fig:velanisotropy}}
\end{figure}

Using these equations, we can build a family of curves giving $v_{\rm
rot,0}/\sigma_0$ as a function of ellipticity for different values of
$\delta$. The result is shown in Figure~\ref{fig:velanisotropy}.  Binney
\& Tremaine (1987) argue that the representative points in the
$(v_{\rm rot}/\sigma, \epsilon)$ plane of systems that are flattened by
rotation lie close to the curve $\delta = 0$, {\em independent} of the
inclination angle $i$.

The mean ellipticity of WLM-1, as measured in
Section~\ref{sec:ellipticity}, is $\epsilon = 0.17$, which is indicated
in Figure~\ref{fig:velanisotropy} by the vertical dashed line.  As the
low mass of the WLM galaxy effectively rules out the possibility that
WLM-1 owes its ellipticity to tidal effects or gravothermal shocks, this
figure very nicely illustrates the kinematical combinations available to
explain the ellipticity of WLM-1.

At one extreme, the ellipticity of WLM-1 could be produced entirely by
rotation.  In this scenario there is no velocity anisotropy
($\delta=0$), and reading from the plot, we see that $v_{\rm rot}/\sigma
\approx 0.57$.  Taking a conservative estimate of 10~\kms\ for the central 
velocity dispersion of the cluster (Sec.~\ref{sec:veldisp}), 
the required rotation would be
5.7~\kms.  However, as we find no measurable rotation ($v_{\rm rot}
\lesssim 0.1$~\kms), we exclude this possibility.

At the other extreme, the ellipticity of WLM-1 is due entirely to
velocity anisotropy.  The quantity $(1-\delta)$, defined as the ratio
between the diagonal terms of the velocity dispersion tensor along the
rotation (``$z$'') axis and along the elongation (``$x$'') axis, should
scale as 

\begin{equation} 
  1-\delta \approx \sigma_z^2/\sigma_x^2.
\end{equation}

\noindent From Figure~\ref{fig:velanisotropy}, one sees that if there is 
zero rotation ($v_{\rm rot}/\sigma=0$), it is possible to produce the
observed ellipticity with a velocity dispersion parameter of $\delta =
0.14$.  Therefore, WLM-1's flattening could be accounted for if the
velocity dispersion along the rotation axis were smaller than along the
elongation axis by
\begin{equation}
\sigma_z \approx 0.93 \, \sigma_x.  
\end{equation}
Again taking the nominal velocity dispersion to be $\sim 10$~\kms, this
velocity anisotropy would only produce a difference of 0.7~\kms\ in
velocity dispersion between the major and minor axes.  Such a difference
would be too small to be detected on the basis of our data. Therefore,
while we cannot prove that velocity anisotropy is responsible for the
observed shape of WLM-1, it certainly appears to be the only
option that is consistent with our observations.


\section{Conclusions} \label{sec:conclusions} 

We have analyzed VLT UVES long-slit spectra and HST WFPC2 images of the
extragalactic globular cluster WLM-1 in an attempt to understand the
cause of its elliptical morphology.  Unlike the situation for Galactic
globular clusters whose detailed shapes may be influenced by their
massive host galaxy, WLM is a low-mass dwarf irregular galaxy,
eliminating the possibilities that the non-sphericity of WLM-1 could be
caused by either tidal stresses or gravothermal shocks.  This leaves two
explanations for the flattening of this cluster: internal rotation
and/or velocity dispersion anisotropy.

Cluster rotation seemed to be a prime candidate for the origin of
WLM-1's ellipticity since two-body relaxation, which erases any
primordial anisotropy, is relatively efficient in star clusters.  In
addition, rotation has been measured for several Galactic globular
clusters using both proper motion and radial velocity techniques [the
review by \citet{Meylan1997} lists the rotation velocities of 11
clusters].  The most notable case is that of $\omega$~Cen, which
exhibits correlated rotation and flattening profiles, strongly
suggesting that its ellipticity is due to rotation.

We therefore obtained high-resolution long-slit echelle spectra along
the major and minor axes of the cluster. We used cross-correlation to
look for velocity shifts on opposite sides of the cluster, dividing the
cluster into annuli in an attempt to obtain a rotation velocity profile.
However, we found no evidence for rotation ($v_{\rm rot} < 0.1$~\kms)
along either the major or minor axes of the cluster.  As the cluster
would require a rotation of $>6$~\kms\ to sustain its shape, we conclude
that rotation is definitely not the main driver of WLM-1's ellipticity.

The only remaining plausible source of flattening in the WLM cluster is
velocity dispersion anisotropy, the main reason elliptical galaxies are
elliptical.  We attempted to measure the stellar velocity dispersion in
WLM-1 based on the width of the cross-correlation peak; however, due to
our large errors, we could see no statistically significant difference
between the velocity dispersions measured along the major and minor
axes.  Calculations reveal that the magnitude of the velocity anisotropy
required to produce the observed ellipticity is relatively small: a 7\%
difference in velocity dispersion, or less than 1~\kms, which is smaller
than our velocity dispersion measurement errors.

In summary, we conclude that the ellipticity of the globular cluster
WLM-1 is most likely due to anisotropy in the stellar velocity
dispersion.

\acknowledgements 

Support for this work was provided by a Princeton-Cat\'olica Prize
Fellowship, Proyecto FONDECYT Regular No.~1030976, and a Gemini Science
Fellowship (AWS), and by Proyecto FONDECYT Regular No.~1030954 (MC).

Based on observations made with ESO Telescopes at the Paranal
Observatories under programme ID 072.D-0355; and with the NASA/ESA 
{\em Hubble Space Telescope}, obtained at the Space Science Institute, 
which is operated by the Association of Universities for Research in 
Astronomy, Inc., under NASA contract NAS5-26555, retrieved from the 
ESO ST-ECF Archive. 



\end{document}